\documentstyle[12pt]{article}

\def\mathfrak{\bf}

\def\be{\begin{equation}}
\def\ee{\end{equation}}
\def\bea{\begin{eqnarray}}
\def\eea{\end{eqnarray}}
\def\dt#1{\on{\hbox{\bf .}}{#1}}                % (big) dot over
\def\Dot#1{\dt{#1}}
\def\IR{\relax{\rm I\kern-.18em R}}
\def\binomial#1#2{\left(\,{\buildrel 
{\raise4pt\hbox{$\displaystyle{#1}$}}\over
{\raise-6pt\hbox{$\displaystyle{#2}$}}}\,\right)}

\def\[{\lfloor{\hskip 0.35pt}\!\!\!\lceil}
\def\]{\rfloor{\hskip 0.35pt}\!\!\!\rceil}

\newcommand{\AmS}{{\protect\the\textfont2
  A\kern-.1667em\lower.5ex\hbox{M}\kern-.125emS}}

\catcode`@=11
\def\un#1{\relax\ifmmode\@@underline#1\else
        $\@@underline{\hbox{#1}}$\relax\fi}
\catcode`@=12

\def\fracm#1#2{\hbox{\large{${\frac{{#1}}{{#2}}}$}}}

\def\ad{{\kern0.5pt
                   \alpha \kern-5.05pt
\raise5.8pt\hbox{$\textstyle.$}\kern
0.5pt}}

\def\Dot#1{{\kern0.5pt
     {#1} \kern-5.05pt \raise5.8pt\hbox{$\textstyle.$}\kern
0.5pt}}

\def\a{\alpha}
\def\b{\beta}
\def\c{\chi}
\def\d{\delta}
\def\e{\epsilon}
\def\f{\phi}
\def\g{\gamma}
\def\h{\eta}

\def\j{\psi}

\def\l{\lambda}

\def\q{\theta}

\def\F{\Phi}

\def\J{\Psi}
\def\L{\Lambda}
\def\O{\Omega}

\def\bo{{\raise.15ex\hbox{\large$\Box$}}}               % D'Alembertian
\def\pa{\partial}                                       % curly d
\def\TH{{\raise.2ex\hbox{$\displaystyle \bigodot$}\mskip-4.7mu \llap H
\;}}
\def\face{{\raise.2ex\hbox{$\displaystyle \bigodot$}\mskip-2.2mu \llap
{$\ddot
        \smile$}}}                                      % happy face
\def\Hat#1{\widehat{#1}}                        % big hat
\def\Bar#1{\overline{#1}}                       % big bar
\def\leftrightarrowfill{$\mathsurround=0pt \mathord\leftarrow \mkern-6mu
        \cleaders\hbox{$\mkern-2mu \mathord- \mkern-2mu$}\hfill
        \mkern-6mu \mathord\rightarrow$}
\def\dvec#1{\vbox{\ialign{##\crcr
        \leftrightarrowfill\crcr\noalign{\kern-1pt\nointerlineskip}
        $\hfil\displaystyle{#1}\hfil$\crcr}}}           % <--> accent
\def\dt#1{{\buildrel {\hbox{\LARGE .}} \over {#1}}}     % dot-over forsp/sb
\def\fracm#1#2{\hbox{\large{${\frac{{#1}}{{#2}}}$}}}
\def\frac#1#2{{\textstyle{#1\over\vphantom2\smash{\raise.20ex
        \hbox{$\scriptstyle{#2}$}}}}}                   % fraction
\def\sfrac#1#2{{\vphantom1\smash{\lower.5ex\hbox{\small$#1$}}\over
        \vphantom1\smash{\raise.4ex\hbox{\small$#2$}}}} % alternatefraction
\def\bfrac#1#2{{\vphantom1\smash{\lower.5ex\hbox{$#1$}}\over
        \vphantom1\smash{\raise.3ex\hbox{$#2$}}}}       % "
\def\afrac#1#2{{\vphantom1\smash{\lower.5ex\hbox{$#1$}}\over#2}}    % "
\def\on#1#2{\mathop{\null#2}\limits^{#1}}               % arbitraryaccent
\newskip\humongous \humongous=0pt plus 1000pt minus 1000pt

\newif\ifdtup

  \def\pp{{\mathchoice
          {%displaystyle
              \kern 1pt%
              \raise 1pt
              \vbox{\hrule width5pt height0.4pt depth0pt
                    \kern -2pt
                    \hbox{\kern 2.3pt
                          \vrule width0.4pt height6pt depth0pt
                          }
                    \kern -2pt
                    \hrule width5pt height0.4pt depth0pt}%
                    \kern 1pt
           }
            {%textstyle
              \kern 1pt%
              \raise 1pt
              \vbox{\hrule width4.3pt height0.4pt depth0pt
                    \kern -1.8pt
                    \hbox{\kern 1.95pt
                          \vrule width0.4pt height5.4pt depth0pt
                          }
                    \kern -1.8pt
                    \hrule width4.3pt height0.4pt depth0pt}%
                    \kern 1pt
            }
            {%scriptstyle
              \kern 0.5pt%
              \raise 1pt
              \vbox{\hrule width4.0pt height0.3pt depth0pt
                    \kern -1.9pt  %[e]=0.15pt
                    \hbox{\kern 1.85pt
                          \vrule width0.3pt height5.7pt depth0pt
                          }
                    \kern -1.9pt
                    \hrule width4.0pt height0.3pt depth0pt}%
                    \kern 0.5pt
            }
            {%scriptscriptstyle
              \kern 0.5pt%
              \raise 1pt
              \vbox{\hrule width3.6pt height0.3pt depth0pt
                    \kern -1.5pt
                    \hbox{\kern 1.65pt
                          \vrule width0.3pt height4.5pt depth0pt
                          }
                    \kern -1.5pt
                    \hrule width3.6pt height0.3pt depth0pt}%
                    \kern 0.5pt%}
            }
        }}

  \def\mm{{\mathchoice
                       {%displaystyle
                             \kern 1pt
               \raise 1pt    \vbox{\hrule width5pt height0.4pt depth0pt
                                  \kern 2pt
                                  \hrule width5pt height0.4pt depth0pt}
                             \kern 1pt}
                       {%textstyle
                            \kern 1pt
               \raise 1pt \vbox{\hrule width4.3pt height0.4pt depth0pt
                                  \kern 1.8pt
                                  \hrule width4.3pt height0.4pt depth0pt}
                             \kern 1pt}
                       {%scriptstyle
                            \kern 0.5pt
               \raise 1pt
                            \vbox{\hrule width4.0pt height0.3pt depth0pt
                                  \kern 1.9pt
                                  \hrule width4.0pt height0.3pt depth0pt}
                            \kern 1pt}
                       {%scriptscriptstyle
                           \kern 0.5pt
             \raise 1pt  \vbox{\hrule width3.6pt height0.3pt depth0pt
                                  \kern 1.5pt
                                  \hrule width3.6pt height0.3pt depth0pt}
                           \kern 0.5pt}
                       }}

\def\pd{{\kern0.5pt
                   + \kern-5.05pt \raise5.8pt\hbox{$\textstyle.$}\kern
0.5pt}}

\def\pmd{{\kern0.5pt
                  \pm \kern-5.05pt \raise6.3pt\hbox{$\textstyle.$}\kern1.5pt}}

\def\md{{\mathchoice
   {%displaystyle
      {{\kern 1pt - \kern-6.2pt \raise5pt\hbox{$\textstyle.$}\kern 1pt}}}
    {%textstyle
      {{\kern 1pt - \kern-6.2pt \raise5pt\hbox{$\textstyle.$}\kern 1pt}}}
    {%scriptstyle
      {\kern0.5pt - \kern-5.05pt \raise3.4pt\hbox{$\textstyle.$}\kern0.5pt}}
    {%scriptscriptstyle
      {\kern0.5pt - \kern-5.05pt \raise3.4pt\hbox{$\textstyle.$}\kern0.5pt}}}}

\def\ad{{\dot{\alpha}}}

\def\pp{{\mathchoice
          {%displaystyle
              \kern 1pt%
              \raise 1pt
              \vbox{\hrule width5pt height0.4pt depth0pt
                    \kern -2pt
                    \hbox{\kern 2.3pt
                          \vrule width0.4pt height6pt depth0pt
                          }
                    \kern -2pt
                    \hrule width5pt height0.4pt depth0pt}%
                    \kern 1pt
           }
            {%textstyle
              \kern 1pt%
              \raise 1pt
              \vbox{\hrule width4.3pt height0.4pt depth0pt
                    \kern -1.8pt
                    \hbox{\kern 1.95pt
                          \vrule width0.4pt height5.4pt depth0pt
                          }
                    \kern -1.8pt
                    \hrule width4.3pt height0.4pt depth0pt}%
                    \kern 1pt
            }
            {%scriptstyle
              \kern 0.5pt%
              \raise 1pt
              \vbox{\hrule width4.0pt height0.3pt depth0pt
                    \kern -1.9pt  %[e]=0.15pt
                    \hbox{\kern 1.85pt
                          \vrule width0.3pt height5.7pt depth0pt
                          }
                    \kern -1.9pt
                    \hrule width4.0pt height0.3pt depth0pt}%
                    \kern 0.5pt
            }
            {%scriptscriptstyle
              \kern 0.5pt%
              \raise 1pt
              \vbox{\hrule width3.6pt height0.3pt depth0pt
                    \kern -1.5pt
                    \hbox{\kern 1.65pt
                          \vrule width0.3pt height4.5pt depth0pt
                          }
                    \kern -1.5pt
                    \hrule width3.6pt height0.3pt depth0pt}%
                    \kern 0.5pt%}
            }
        }}

  \def\mm{{\mathchoice
                       {%displaystyle
                             \kern 1pt
               \raise 1pt    \vbox{\hrule width5pt height0.4pt depth0pt
                                  \kern 2pt
                                  \hrule width5pt height0.4pt depth0pt}
                             \kern 1pt}
                       {%textstyle
                            \kern 1pt
               \raise 1pt \vbox{\hrule width4.3pt height0.4pt depth0pt
                                  \kern 1.8pt
                                  \hrule width4.3pt height0.4pt depth0pt}
                             \kern 1pt}
                       {%scriptstyle
                            \kern 0.5pt
               \raise 1pt
                            \vbox{\hrule width4.0pt height0.3pt depth0pt
                                  \kern 1.9pt
                                  \hrule width4.0pt height0.3pt depth0pt}
                            \kern 1pt}
                       {%scriptscriptstyle
                           \kern 0.5pt
             \raise 1pt  \vbox{\hrule width3.6pt height0.3pt depth0pt
                                  \kern 1.5pt
                                  \hrule width3.6pt height0.3pt depth0pt}
                           \kern 0.5pt}
                       }}

\def\pd{{\kern0.5pt
                   + \kern-5.05pt \raise5.8pt\hbox{$\textstyle.$}\kern
0.5pt}}

\def\pmd{{\kern0.5pt
                  \pm \kern-5.05pt \raise6.3pt\hbox{$\textstyle.$}\kern1.5pt}}

\def\md{{\mathchoice
   {%displaystyle
      {{\kern 1pt - \kern-6.2pt \raise5pt\hbox{$\textstyle.$}\kern 1pt}}}
    {%textstyle
      {{\kern 1pt - \kern-6.2pt \raise5pt\hbox{$\textstyle.$}\kern 1pt}}}
    {%scriptstyle
      {\kern0.5pt - \kern-5.05pt \raise3.4pt\hbox{$\textstyle.$}\kern0.5pt}}
    {%scriptscriptstyle
      {\kern0.5pt - \kern-5.05pt \raise3.4pt\hbox{$\textstyle.$}\kern0.5pt}}}}

\def\dslash{\not{\hbox{\kern-2pt $\partial$}}}
\def\Dslash{\not{\hbox{\kern-4pt $D$}}}
\def\pslash{\not{\hbox{\kern-2.3pt $p$}}}
 \newtoks\slashfraction
 \slashfraction={.13}
 \def\slash#1{\setbox0\hbox{$ #1 $}
 \setbox0\hbox to \the\slashfraction\wd0{\hss \box0}/\box0 }
 
\font\ro=cmsy10                          % font with rope
\def\kcr{{\hbox{\ro \char'170}}}                % right-handed rope
\def\ktl{{\hbox{\ro \char'170}}}        % top end for left-handed rope
\def\ktr{{\hbox{\ro \char'170}}}        % " right
\def\kbl{{\hbox{\ro \char'170}}}        % " bottom left
\def\kbr{{\hbox{\ro \char'170}}}        % " right
\def\plpl{\raise-2pt\hbox{$\raise3pt\hbox{$_+$}\hskip-6.67pt\raise0.0pt
\hbox{$^+$}\hskip 0.01pt$}}
\def\mimi{\raise-2pt\hbox{$\raise3pt\hbox{$_-$}\hskip-6.67pt\raise0.0pt
\hbox{$^-$}\hskip 0.01pt$}} 

\def\bo{{\raise.15ex\hbox{\large$\Box$}}}               % D'Alembertian
\def\pa{\partial}                                       % curly d
\def\TH{{\raise.2ex\hbox{$\displaystyle \bigodot$}\mskip-4.7mu \llap H \;}}
\def\face{{\raise.2ex\hbox{$\displaystyle \bigodot$}\mskip-2.2mu \llap {$\ddot
        \smile$}}}                                      % happy face
\def\Hat#1{\widehat{#1}}                        % big hat
\def\Bar#1{\overline{#1}}                       % big bar
\def\leftrightarrowfill{$\mathsurround=0pt \mathord\leftarrow \mkern-6mu
        \cleaders\hbox{$\mkern-2mu \mathord- \mkern-2mu$}\hfill
        \mkern-6mu \mathord\rightarrow$}
\def\dvec#1{\vbox{\ialign{##\crcr
        \leftrightarrowfill\crcr\noalign{\kern-1pt\nointerlineskip}
        $\hfil\displaystyle{#1}\hfil$\crcr}}}           % <--> accent
\def\dt#1{{\buildrel {\hbox{\LARGE .}} \over {#1}}}     % dot-over for sp/sb
\def\fracm#1#2{\hbox{\large{${\frac{{#1}}{{#2}}}$}}}
\def\frac#1#2{{\textstyle{#1\over\vphantom2\smash{\raise.20ex
        \hbox{$\scriptstyle{#2}$}}}}}                   % fraction
\def\sfrac#1#2{{\vphantom1\smash{\lower.5ex\hbox{\small$#1$}}\over
        \vphantom1\smash{\raise.4ex\hbox{\small$#2$}}}} % alternate fraction
\def\bfrac#1#2{{\vphantom1\smash{\lower.5ex\hbox{$#1$}}\over
        \vphantom1\smash{\raise.3ex\hbox{$#2$}}}}       % "
\def\afrac#1#2{{\vphantom1\smash{\lower.5ex\hbox{$#1$}}\over#2}}    % "
\def\on#1#2{\mathop{\null#2}\limits^{#1}}               % arbitrary accent
\parskip=\medskipamount                 % space between paragraphs (LaTeX)
\thicklines                         % thick straight lines for pictures (LaTeX)

\def\oldheadpic{                                % old UM heading
        \setlength{\unitlength}{.4mm}
        \thinlines
        \par
        \begin{picture}(349,16)
        \put(325,16){\line(1,0){4}}
        \put(330,16){\line(1,0){4}}
        \put(340,16){\line(1,0){4}}
        \put(335,0){\line(1,0){4}}
        \put(340,0){\line(1,0){4}}
        \put(345,0){\line(1,0){4}}
        \put(329,0){\line(0,1){16}}
        \put(330,0){\line(0,1){16}}
        \put(339,0){\line(0,1){16}}
        \put(340,0){\line(0,1){16}}
        \put(344,0){\line(0,1){16}}
        \put(345,0){\line(0,1){16}}
        \put(329,16){\oval(8,32)[bl]}
        \put(330,16){\oval(8,32)[br]}
        \put(339,0){\oval(8,32)[tl]}
        \put(345,0){\oval(8,32)[tr]}
        \end{picture}
        \par
        \thicklines
        \vskip.2in}
\def\oldtitle#1#2#3#4{\oldheadpic\begin{center}\vglue.5in{\large\bf #1}\\[.6in]
        {#2}\\[.1in] {\it Department of Physics and Astronomy}\\
        {\it University of Maryland, College Park, MD 20742}\\[.6in]
        Physics Publication \#{#3}\\ {#4}\\[1.5in] {\bf ABSTRACT}\\[.1in]
        \end{center} \begin{quotation}}                 % old title stuff
\def\oldTitle#1#2#3#4#5#6#7{\oldheadpic\begin{center} \vglue .4in
        {\large\bf #1}\\[.4in]
        {#2}\\[.1in] {\it Department of Physics and Astronomy}\\
        {\it University of Maryland, College Park, MD 20742}\\[.1in]
        {#3}\\[.1in] {\it {#4}}\\ {\it {#5}}\\[.4in]
        Physics Publication \#{#6}\\ {#7}\\[.5in] {\bf ABSTRACT}\\[.1in]
        \end{center} \begin{quotation}}                 % " for 2 authors
\def\border{                                            % border
        \setlength{\unitlength}{1mm}
        \newcount\xco
        \newcount\yco
        \xco=-21
        \yco=12
        \begin{picture}(140,0)
        \put(\xco,\yco){$\ktl$}
        \advance\yco by-1
        {\loop
        \put(\xco,\yco){$\kcr$}
        \advance\yco by-2
        \ifnum\yco>-240
        \repeat
        \put(\xco,\yco){$\kbl$}}
        \xco=158
        \yco=12
        \put(\xco,\yco){$\ktr$}
        \advance\yco by-1
        {\loop
        \put(\xco,\yco){$\kcr$}
        \advance\yco by-2
        \ifnum\yco>-240
        \repeat
        \put(\xco,\yco){$\kbr$}}
        \put(-20,13){\tiny University of Maryland Elementary Particle
Physics University of Maryland Elementary Particle Physics University of
Maryland Elementary Particle Physics}
        \put(-20,-241.5){\tiny University of Maryland Elementary
Particle Physics University of Maryland Elementary Particle Physics
University of Maryland Elementary Particle Physics}
        \end{picture}
        \par\vskip-8mm}
\def\bordero{                                           % alternate border
        \setlength{\unitlength}{1mm}
        \newcount\xco
        \newcount\yco
        \xco=-31
        \yco=12
        \begin{picture}(140,0)
        \put(\xco,\yco){$\ktl$}
        \advance\yco by-1
        {\loop
        \put(\xco,\yco){$\kclr}
        \advance\yco by-2
        \ifnum\yco>-240
        \repeat
        \put(\xco,\yco){$\kbl$}}
        \xco=151
        \yco=12
        \put(\xco,\yco){$\ktr$}
        \advance\yco by-1
        {\loop
        \put(\xco,\yco){$\kcr$}
        \advance\yco by-2
        \ifnum\yco>-240
        \repeat
        \put(\xco,\yco){$\kbr$}}
        \put(-20,12){\ooo bacdefghidfghghdhededbihdgdfdfhhdheidhdhebaaahjhhdahba

hgdedge
   hgfdiehhgdigicba}
        \put(-20,-241.5){\ooo ababaighefdbfghgeahgdfgafagihdidihiidhiagfedhadbfd

ecdcdfa
   gdcbhaddhbgfchbgfdacfediacbabab}
        \end{picture}
        \par\vskip-8mm}
\def\headpic{                                           % UM heading
        \indent
        \setlength{\unitlength}{.4mm}
        \thinlines
        \par
        \begin{picture}(29,16)
        \put(165,16){\line(1,0){4}}
        \put(170,16){\line(1,0){4}}
        \put(180,16){\line(1,0){4}}
        \put(175,0){\line(1,0){4}}
        \put(180,0){\line(1,0){4}}
        \put(185,0){\line(1,0){4}}
        \put(169,0){\line(0,1){16}}
        \put(170,0){\line(0,1){16}}
        \put(179,0){\line(0,1){16}}
        \put(180,0){\line(0,1){16}}
        \put(184,0){\line(0,1){16}}
        \put(185,0){\line(0,1){16}}
        \put(169,16){\oval(8,32)[bl]}
        \put(170,16){\oval(8,32)[br]}
        \put(179,0){\oval(8,32)[tl]}
        \put(185,0){\oval(8,32)[tr]}
        \end{picture}
        \par\vskip-6.5mm
        \thicklines}
\def\title#1#2#3#4{\border\headpic {\hbox to\hsize{#4 \hfill UMDEPP #3}}\par
        \begin{center} \vglue .5in {\large\bf #1}\\[.6in]
        {#2}\\[.1in] {\it Department of Physics and Astronomy}\\
        {\it University of Maryland, College Park, MD 20742}\\[1.5in]
        {\bf ABSTRACT}\\[.1in] \end{center} \begin{quotation}}  % title stuff
\def\Title#1#2#3#4#5#6#7{\border\headpic
        {\hbox to\hsize{#7 \hfill UMDEPP #6}}\par
        \begin{center} \vglue .4in {\large\bf #1}\\[.4in]
        {#2}\\[.1in] {\it Department of Physics and Astronomy}\\
        {\it University of Maryland, College Park, MD 20742}\\[.1in]
        {#3}\\[.1in] {\it {#4}}\\ {\it {#5}}\\[.5in] {\bf ABSTRACT}\\[.1in]
        \end{center} \begin{quotation}}                 % " for 2 authors
\def\endtitle{\end{quotation}\newpage}                  % end title page

\def\qd{{\kern0.5pt
                   q \kern-5.05pt \raise5.8pt\hbox{$\textstyle.$}\kern
0.5pt}}

\begin{document}

\def\dt#1{\on{\hbox{\bf .}}{#1}}                % (big) dot over
\def\Dot#1{\dt{#1}}

\def\gfrac#1#2{\frac {\scriptstyle{#1}}
        {\mbox{\raisebox{-.6ex}{$\scriptstyle{#2}$}}}}
\def\gg{{\hbox{\sc g}}}
\border\headpic {\hbox to\hsize{August 2002 \hfill
{UMDEPP 03-003}}}
\par
\par
\setlength{\oddsidemargin}{0.3in}
\setlength{\evensidemargin}{-0.3in}
\begin{center}
\vglue .10in
{\large\bf Dynamical Superfield Theory of Free Massive Superspin-1 Multiplet}
\\[.5in]
I.L. Buchbinder$^a$\footnote{joseph@tspu.edu.ru}, S. James Gates, Jr.$^b$\footnote{gatess@wam.umd.edu}, W. D. Linch,
III$^b$\footnote{ldw@physics.umd.edu} and J. 
Phillips$^b$\footnote{ferrigno@physics.umd.edu}
\\[0.05in]
$^a${\it Department of Theoretical Physics, Tomsk State Pedagogical
University\\ 634041 Tomsk, Russia}
\\[0.05in]
$^b${\it Department of Physics, University of Maryland\\ 
College Park, MD 20742-4111 USA}

{\bf ABSTRACT}\\[.01in]
\end{center}
\begin{quotation}
{We construct an $N=1$ supersymmetric Lagrangian model for the massive
superspin-1 superfield.  The model is described by a dynamical spinor
superfield and an auxiliary chiral scalar superfield.  On-shell this
model leads to a spin-${\fracm 12}$, spin-${\fracm 32}$ and two spin-1
propagating component fields.   The superfield action is given and its
structure in the fermionic component sector is presented.  We prove that
the most general theory is characterized by a one parameter family of
actions.  The massless limit is shown to correspond to the dynamics of
both the gravitino and superhelicity-${\fracm 12}$ multiplets.}

${~~~}$ \newline

\endtitle
\noindent
%%%%%%%%%%%%%%%%%%%%%%%%%%%%%%%%%%%%%%%%%%%%%%%%
{\bf 1.}~~Construction of a supersymmetric Lagrangian formulation for
free arbitrarily high spin massive fields is still an unsolved problem 
in field theory.  Although free non-supersymmetric Lagrangian models 
for any integer or half-integer spin massive field have been presented 
long ago \cite{Singh}, the off-shell supersymmetrization of these theories
is  a non-trivial problem.  For this purpose the results of ref.\
\cite{Singh}, given in the conventional field theory formalism, are
practically useless.  This is to be expected since these theories have a
complicated auxiliary field structure, which would lead to complicated
off-shell supersymmetry transformations.   Therefore, we must develop a
new, quite independent approach.  The realization of supersymmetric
theories is usually carried out in the framework of the superfield
formalism.  Superfield methods were the basis for solving the analogous
problem for supersymmetric massless theories \cite{Kuzenko} (see also
\cite{SJGWS,Buch1}) both for Minkowski and for AdS spaces.

In the recent paper \cite{Buch2} we have constructed a superfield 
Lagrangian model for the $N=1$ massive multiplet with superspin-${\fracm 
32}$ having the highest spin-$2$.  This model is a natural off-shell
supersymmetric generalization of the known Fierz-Pauli theory \cite{Fierz}
and describes the dynamics of  massive component fields with the spins $2,
{\fracm 32}$ and $1$.  We also found that such a superfield formulation
demands taking into account an auxiliary general vector superfield.\footnote
{See the on-shell formulation of the dynamical superspin-${\fracm 32}$ multiplet in a
recent preprint \cite{Zino1}.}

The purpose of this letter is to continue the work \cite{Buch2} and develop 
a superfield Lagrangian formulation describing the dynamics of a massive
multiplet with superspin-$1$ having the highest component field of spin-$3/2$. 
Such a model looks like a supersymmetric generalization of the massive
Rarita-Schwinger theory \cite{Rarita}.  It is worth noting from the beginning
that the structure of the models with integer superspin superfields should 
differ from theories with half integer superspin.  This is analogous to the
difference between conventional field models with integer and half integer 
spin fields \cite{Singh}.

To derive a superfield action for the massive superspin-$1$ model we will 
refer to the procedure outlined in ref \cite{Buch2}. We begin with a superfield 
that carries the massive irreducible representation of the Poincar\'e
supergroup.  The corresponding representations are characterized by the mass 
$m$ and superspin $Y$.  On-shell, they contain propagating component fields 
of spins $(Y-1/2, Y, Y, Y+1/2)$ \cite{Buch1}. Using a suitable superfield, we
construct the most general quadratic superfield action.  This action reproduces
the conditions that describe the irreducible representations in the space of
superfields \cite{Buch1} as a concequence of the equations of motion.  This
procedure fails if we work with only the superfields corresponding to the given
irreducible representation(we call such superfields physical).  The solution to
this problem is to couple the physical superfield to an auxiliary superfield
within the action.  In the case of superspin-$1$, the role of the physical
superfield is played by a complex spinor superfield $V_{\alpha}$ and the
auxiliary superfield is a chiral scalar superfield ${\Phi}$.

We  note activity and recent progress in higher spin field theories.  It was noted 
in ref \cite{Biswas} that massive irreducible representations in Minkowski and 
(A)dS spaces can, in principle, be obtained from massless theories by dimensional 
reduction.  This suggests a derivation of  massive models from massless models 
in higher dimensions.   Higher spin massless models have been formulated in 
arbitrary dimensional constant curvature spaces \cite{Lopatin}.\footnote{Also it 
is worth pointing out a new development of massless higher spin theories
\cite{Francia}\cite{Segal}.}  This may open a possibility for constructing massive  models
in constant curvature spaces.   However, applying the above described procedure to
supersymmetic theories can not be straightforward since supersymmetry  has not been
universally formulated for all dimensions.  Alternatively, supersymmetric  massive higher
spin field theories may also be derived from superstring theory  since any string model
contains an infinite tower of higher spin massive modes (see e.g.
\cite{Berk}).  However, there is no guarantee that such an approach leads to 
field dynamics corresponding to irreducible representations of the Poincar\'e
supergroup. Finally, there was some progress in the study of massive field
dynamics in constant curvature space \cite{Buch3} - \cite{Zino2}.  The
corresponding supersymmetric formulation is unknown.

This letter is organized as follows. We discuss the conditions necessary
for an arbitrary superfield to form an irreducible massive superspin-$1$
representation. Then, we construct the superfield action that leads to
these conditions as a consequence of the equations of motion. To
clarify the structure of the superfield action obtained, we analyze the
form of this action in the fermionic sector and show that it reproduces 
the conventional Rarita-Schwinger and Dirac on-shell equations. We give a
brief discussion of the massless limit and find that it is equivalent
to standard superfield actions for the gravitino \cite{SJGWS} and the 
superhelicity-${\fracm  12}$ supermultiplets \cite{Buch1}.

%%%%%%%%%%%%%%%%%%%%%%%%%%%%%%%%%%%%%%%%%%%%%%%%
\noindent
{\bf 2.}~~~~We seek a multiplet which contains a massive spin-3/2 field at 
the $\q\bar\q$ level.  We would also like this theory to be comparable with
the known gravitino multiplet in the massless limit.  Since the gravitino
multipet is descibed by a spinor superfield \cite{SJGWS}, we choose an arbitrary
spinor superfield $V_\a$ as the physical superfield.  We must find the
appropriate conditions for this field to form an irreducible representation of
massive supersymmetry.  To find these conditions we will use the theory of projectors
developed in \cite{Siegel1} and subsequently modified for massive superfields
(see, for example, \cite{Buch1}).  First, this field must satisfy the Dirac 
equation\footnote{In this equation and throughout this presentation, we use 
{\it Superspace}\cite{Gates1} conventions.  Here an underlined vector index
simultaneously denotes the usual Minkowski 4-vector index as well as a pair of
undotted and dotted Weyl spinor indices.}
\bea
\label{Dirac1/2}
i\pa_{\un a}\Bar V^{\dot\a}+mV_\a=0
\eea
Next, we seek the conditions that diagonalize the superspin casimir operator.   The
superspin operator, as described in \cite{Buch1}, acting on this field is given by:
\bea
{\bf C}V_\a=m^4({\fracm 34}I+{\fracm 34}{\cal P}_{(0)}+{\bf B})V_\a
\eea
Where ${\cal P}_{(0)}$ is the linear subspace projection operator and ${\bf B}$ is given by:
\bea
{\bf B}={1\over 4m^2}(M_{\a\b}{\bf P}^\b_{~\dot\a}-\Bar M_{\dot\a\dot\b}{\bf P}_\a^{~\dot\b})[D^\a,\Bar D^{\dot\a}]
\eea
Here, ${\bf P}_{\un a}=-i\pa_{\un a}$, the momentum operator, and $M_{\a\b}$ is the Lorentz
generator written in the spinor representation $SL(2,C)$.  If $V_\a$ is in the chiral subspace, it would
satisfy a superspin-1/2 representation since ${\cal P}_{(0)}V_\a=0$ and ${\bf B} V_\a=0$.  If $V_\a$ is
in the linear subspace, i.e.
\bea
\label{subspace}
D^2V_\a=0~;~\Bar D^2V_\a=0
\eea
then the combination ${\bf B}V_\a$ becomes:
\bea
\label{BV}
{\bf B}V_\a = {1\over 8m^2}D_\a\Bar D^2D_\b V^\b+{\fracm 12}V_\a
\eea
If $V_\a$ satisifies the following conditions:
\bea
\label{irrep}
D^\a V_\a=0~;~\Bar D_{\dot\a} \Bar V^{\dot\a}=0
\eea
then ${\bf B}V_\a={\fracm 12}V_\a$.  Thus ${\bf C}V_\a=m^42V_\a=m^41(1+1)V_\a$, and
$V_\a$ satisfies a superspin-$1$ representation.  At the component level this representation
contains spin-$({\fracm 12}, 1, 1, {\fracm 32})$ fields.  The conditions (\ref {subspace})
restricts $V_{\a}$ to the linear subspace, while the conditions (\ref {irrep}) select out
the superspin-1 state of that subspace.  Note, that because this field has only one
species of spinor index, we do not need the supplementary condition
$\pa_{\un a}V^{\a}=0$ which is usually applied to massive representations.  

Thus, we have shown that an irreducible superspin-1 representation can be obtained using
the general superfield $V_\a$.  At the $\q\bar\q$ level, this representation contains a
spin-${\fracm 32}$ component field.  The superfield $V_\a$ must satisfy the Dirac
equation (\ref{Dirac1/2}), and be in the linear subspace (\ref{subspace}).  We also
require the conditions (\ref{irrep}) to select the superspin-1 state of the linear
subspace.

%%%%%%%%%%%%%%%%%%%%%%%%%%%%%%%%%%%%%%%%%%%%%%%%%%%%%%%%%%%%%%%%%%%%%%%%%%%%%%%%%%%%
\noindent
{\bf 3.}~~~~Now that we know the conditions required to make $V_\a$ a superspin-$1$
irreducible representation, we search for an action that reproduces these conditions as a
consequence of the equations of motion.  We proceed by writing the most general action
quadratic in $V_\a$.  Then, we show that this action can not produce the required on-shell
equations;(\ref{Dirac1/2}), (\ref{subspace}), (\ref{irrep}).  Finally, we show that the
superspin-$1$ representation can be obtained by coupling $V_\a$ to a chiral scalar
superfield $\F$.

To begin constructing the action, we set the mass dimension of $V_\a$ so that the spin-3/2
component field has the canonical mass dimension.  The most general action quadratic in
$V_\a$ and constructed from spinorial covariant derivatives is:
\bea
{\cal S}[V_\a]=\int d^8z\Big\{\a_1V^\a D_\a \Bar D_{\dot\a}\Bar V^{\dot \a}
+\a_2V^\a \Bar D_{\dot\a} D_\a \Bar V^{\dot\a}
+m(V^\a V_\a+\Bar V_{\dot\a} \Bar V^{\dot\a})\cr
+\b V^\a D^2V_\a +\b^*\Bar V_{\dot\a}\Bar D^2 \Bar V^{\dot\a}
+\g V^\a\Bar D^2 V_\a +\g^*\Bar V_{\dot\a}D^2 \Bar V^{\dot\a} \Big\}
\eea
Here $\a_1$ and $\a_2$ are real.  The equation of motion for the superfield $V_\a$ is:
\bea
\label{firsteom}
E_\a :={\d{\cal S}\over \d V^\a}=\a_1 D_\a \Bar D_{\dot\a} \Bar V^{\dot\a}
+\a_2 \Bar D_{\dot\a} D_\a \Bar V^{\dot\a}
+2m V_\a
+2\b D^2 V_\a
+2\g \Bar D^2 V_\a=0
\eea
Taking $\Bar D^2 E_\a=0$ and setting $\b=\a_1=0$, implies that $\Bar D^2 V_\a=0$. 
The equation of motion now takes the form:
\bea
\label{eom2}
E_\a =\a_2 \Bar D_{\dot\a} D_\a \Bar V^{\dot\a}
+2m V_\a=0
\eea
Next, taking $D^\a E_\a=0$ yields:
\bea
\label{badnews}
-{\a_2 \over 2}D^2 \Bar D_{\dot\a} \Bar V^{\dot\a}
+2m D^\a V_\a=0
\eea
and we are forced to set $\a_2=0$ if we want $D^\a V_\a=0$.  With this choice,
(\ref{eom2}) now reads $2mV_\a=0$, thus elliminating the entire superfield $V_\a$
on-shell.   This procedure fails whether we choose to set $\Bar D^2 E_\a=0$ first, as in
this derivation, or
$D^\a E_\a=0$ first.  This means that the most general action can not produce the proper
on-shell equations to make $V_\a$ an irreducible representation.  This does not come as a
surprise.  In ref \cite{SJGWS}, it was shown that the most general {\it massless} action
can not give rise to a single irreducible representation of supersymmetry\cite{Siegel1}. 
Using the same arguments as in \cite{SJGWS}, one can see that the addition of mass terms
alone can not ensure that the action describes one irreducible representation.  We need a
different mechanism to remove the unwanted subspaces and make $V_\a$ the proper irreducible
representation.

We are forced to couple $V_\a$ to an auxiliary field to alleviate the situation.  If the
auxiliary field vanishes on-shell, the consistency of the equation of motion will then imply
a differential constraint on $V_\a$.  Once $V_\a$ is constrained, the above action can
reproduce the correct on-shell conditions.  We choose a chiral scalar superfield $\F$ with
the hope of removing the first term in equation (\ref{badnews}).  This term arises from
the variation of a coupling term of the form $\F \Bar D^{\dot\a} \Bar V_{\dot\a}$.  The most
general form of the auxillary sector of the action is the following:
\bea
{\cal S}_{aux}[V_\a ,\F]=
\int d^8z \big\{ -{\fracm 12} \F D^\a V_\a 
-{\fracm 12} \Bar\F \Bar D_{\dot\a}\Bar V^{\dot\a}
+\g_1\F\Bar\F \Big\}\cr
+{m\over2}\g_2\int d^6z \F\F
+{m\over2}\g_2^*\int d^6\bar z \Bar\F\Bar\F 
\eea
The equations of motion become:
\bea
\label{neomV}
{\d ({\cal S}+{\cal S}_{aux}) \over {\d V^\a}}=0\Rightarrow
E_\a
+{\fracm 12}D_\a\F=0
\eea
and for the auxiliary field:
\bea
\label{aux1}
{\d {\cal S}_{aux}\over \d\F}=0\Rightarrow
+{\fracm 18}\Bar D^2 D^\a V_\a
-{\fracm 14}\g_1\Bar D^2\Bar\F
+m\g_2\F=0
\eea
Note that if $\F =0$ we would have $\Bar D^2D^\a V_\a =0$, the desired differential constraint.  Contracting
$\Bar D^2 D^\a$ on (\ref{neomV}) we have:
\bea
\g_3\Bar D^2 D^2 \Bar D_{\dot\a}\Bar V^{\dot\a}
+2m \Bar D^2 D^\a V_\a
+8\Box \F=0
\eea
where $\g_3=\a_1-{\fracm 12}\a_2$.  
Using the equation of motion for $\F$ we have:
\bea
\Big\{ 1+4\g_1\g_3 \Big\}8\Box\F
+\Big \{ \g_1-2\g_2^*\g_3 \Big \} 4 m \Bar D^2 \Bar \F
-16m^2\g_2 \F=0
\eea
The following choices of coefficients will constrain $\F$ to vanish:
\bea
\g_1=-{1\over {4\g_3}}~;~\g_2=-{1\over {8{\g_3}^2}}
\eea
From (\ref{aux1}), the vanishing of $\F$ implies the following condition on $V_\a$:
\bea
\Bar D^2 D_\a V^\a=0~;~D^2\Bar D_{\dot\a}\Bar V^{\dot\a}=0
\eea
The equation of motion for $V_\a$, now takes the same form as equation (\ref{firsteom}).  Multiplying by
$\Bar D^2$ and setting $\a_1=\b=0$ we have $\Bar D^2 V_\a=0$, and $D^2 \Bar V_{\dot\a}=0$. 
Then the equation of motion becomes (\ref{eom2}) exactly.  Contracting $D^\a$ on equation
(\ref{eom2}) now yields $2m D^\a V_\a=0$.  With this result, $V_\a$ is fully irreducible
in the superspace.  The equation of motion now takes the form of the Dirac equation:
\bea
-2i\a_2 \pa_{\un a}\Bar V^{\dot\a}+2mV_\a=0
\eea
Thus, with $\a_2=-1$, $V_\a$ will satisfy (\ref{Dirac1/2}), (\ref{subspace}), and
(\ref{irrep}) on-shell.  We can now write the full action:
\bea
\label{finalaction}
{\cal S}[V_\a,\F]=\int d^8z\Big \{
-V^\a\Bar D_{\dot\a}D_\a\Bar V^{\dot\a}
+m(V^\a V_\a +\Bar V_{\dot\a}\Bar V^{\dot\a})
+\g V^\a\Bar D^2V_\a~~~~~~~~~\cr
+{\g}^*\Bar V_{\dot\a} D^2\Bar V^{\dot\a}
-{\fracm 12}\F\Bar\F
-{\fracm 12}\F D^\a V_\a-{\fracm 12}\Bar \F \Bar D_{\dot\a} \Bar V^{\dot\a}
 \Big\}
-{\fracm m4} \int d^6z\F\F-{\fracm m4}\int d^6 \bar z \Bar\F\Bar\F
\eea
We would like to point out that all coefficients have been determined except for
$\g$.  The $\g$ terms are interesting because they are purely auxiliary.  We illustrate
this by formally integrating over the superspace, i.e.
\bea
\label{zero}
\int d^8z \Big\{\g V^\a\Bar D^2V_\a\Big\}=
{\fracm 18} \int d^4x \Big\{ \g D^2\Bar D^2V^\a |\Bar D^2 V_\a |
-{\fracm {\g}2}D^\b\Bar D^2V^\a |D_\b\Bar D^2V_\a |\Big\}
\eea
On-shell (\ref{zero}) vanishes since $\Bar D^2V_\a=D^2V_\a=0$.  We can also exhibit the
irrelevance of these terms by considering the following field redefinition:
\bea
V_\a \rightarrow V_\a+{a\over m}\Bar D^2 V_\a
\eea
where $a$ is an arbitrary complex number.  With this redefinition we can change
$\g$ to any arbitrary number, without influencing the on-shell results of the
Lagrangian.  Although these terms are purely auxiliary in the massive
theory, they will play an important role in understanding the massless limit.  

We have given a two real parameter family of actions, governed by one complex parameter.  These
actions lead to the on-shell equations which describe an irreducible superspin-1 multiplet. 
The off-shell degrees of freedom include the physical spinor superfield $V_\a$ and an
auxiliary chiral scalar field $\F$.  On-shell $\F=0$, and $V_\a$ becomes an irreducible
representation of the Poincar\'e supergroup having superspin-1.

%%%%%%%%%%%%%%%%%%%%%%%%%%%%%%%%%%%%%%%%%%%%%%%
\noindent
{\bf 4}.  Here we give the fermionic component action and show how the equations of motion
reproduce the correct fermionic massive representations.  We will see that there are two
propagating fermions of spin-${\fracm 12}$ and spin-${\fracm 32}$.  The spin-${\fracm 32}$
field has no divergence and satisfies the Rarita-Schwinger equation, while the
spin-${\fracm 12}$ field satisfies the Dirac equation.

Using the following component definitions:
\bea
V_\a | = \l_\a~&;&~D_\a\F |=\f_\a\cr
-{\fracm 14}D^2V_\a | = \h_\a~&;&~
-{\fracm 14}{\bar D^2}V_\a | = \c_a\cr
D_\a \bar D_{\dot\a} V_\b| = \j_{\a\dot\a\b}~&;&~
{\fracm 1{16}}D^2 {\bar D^2}V_\a| = \L_\a
\eea
we formally integrate over the fermionic coordinates.  The fermionic sector of the
action becomes:
\bea
{\cal S}_f=\int d^4x \Big \{\j^{\b\un a}({\fracm i4}\pa_{\un b}\Bar\j^{\dot\b}_{~\un a}
+{\fracm m4}\j_{\b\un a}
-{\fracm i4}\pa_{\un a}\f_\b)\cr
+{\fracm i8}\f^{\a}\pa_{\un a}\Bar\f^{\dot\a}
+{\fracm m8}\f^\a\f_\a
+{\fracm 12}\f^\a\L_\a
+\L^\a\Bar\j_{\dot\a\a}^{~~\a}\cr
+2m\L^\a\l_\a
-8\g\c^\a\L_\a 
+i\c^\a\pa_{\un a}\Bar\c^{\dot\a}
+2m\c^\a\h_\a
+ h.c.\Big\}
\eea
It is trivial to see that on-shell $\L_\a=\c_\a=\h_\a=0$.  By taking the divergence of the $\Bar\j$
equation of motion, such that the $\Bar\f$ term becomes $i\Box\Bar\f^{\dot\a}$, and substituting the $\f$
equation of motion, we see that $\f_\a=0$.  This leaves the following equations of motion:
\bea
\label{psi-eom}
i\pa_{\un b}\Bar\j^{\dot\b}_{~~\un a}
+m\j_{\b\un a}=0
\eea
\bea
\label{phi-eom}
\pa^{\un b}\j_{\a\un b}=0
\eea
\bea
\label{Lambda-eom}
\Bar\j_{\un a}^{~~\dot\a}+2m\l_{\a}=0
\eea
The trace of (\ref{psi-eom}) - (\ref{phi-eom}) yields the Dirac equation on
$\j^\a_{~\dot\a\a}$:
\bea
i \pa_{\un a}\j^{\b\dot\a}_{~~\b} +m\Bar \j^{\dot\b}_{~\a\dot\b}=0
\eea
Thus, (\ref{Lambda-eom}) implies that $\l_\a$ also satisfies the Dirac equation.  Using
these facts, one can show that the trace of (\ref{psi-eom})+(\ref{phi-eom}) leads to:
\bea
\label{nodiv}
\pa^{\un b}\Bar\j_{(\dot\a\dot\b)\b}+2i\Box\Bar\l_{\dot\a}=0\cr
\Rightarrow\pa^{\un b}(\Bar\j_{(\dot\a\dot\b)\b}-{\fracm 23}i\pa_{\b(\dot\a}\Bar\l_{\dot\b)})=0
\eea
Taking this symmetric divergenceless field as the gravitino:
\bea
\Hat{\Bar{\bf\J}}_{\dot\a\dot\b\b}:=\Bar\j_{(\dot\a\dot\b)\b}-{\fracm
23}i\pa_{\b(\dot\a}\Bar\l_{\dot\b)}
\eea
we can now show that the symmetric part of (\ref{psi-eom}) implies the Rarita-Schwinger
equation on $\Hat{\bf\J}$.
\bea
&~&-{\fracm i4}\pa^{\dot\b}_{~(\b}\Bar\j_{\a)(\dot\a\dot\b)}
+{\fracm i4}\pa_{\dot\a(\b}\Bar\j^{\dot\g}_{~\a)\dot\g}
+{\fracm m2}\j_{(\a\b)\dot\a}\cr
&=&-{\fracm i4}\pa^{\dot\b}_{~(\b}\Hat{\Bar{\bf\J}}_{\a)\dot\a\dot\b}
+{\fracm m2}\Hat{\bf\J}_{\a\b\dot\a}\cr
&+&(-{\fracm 16}+{\fracm 12}-{\fracm 13})\pa_{\dot\a(\b}\pa_{\a)\dot\b}\Bar\l^{\dot\b}=0
\eea
Here, we have used (\ref{Lambda-eom}) and the Dirac equation on $\Bar\l_{\dot\b}$
extensively.  Thus, the gravitino satisfies the Rarita-Schwinger equation:
\bea
\label{Dirac32}
-{\fracm i2}\pa^{\dot\b}_{~(\b}\Hat{\Bar{\bf\J}}_{\a)\dot\a\dot\b}
+m\Hat{\bf\J}_{\a\b\dot\a}
={\fracm 12}\e_{\un a \un b cd}\pa^{[c}\Hat{\Bar{\bf\J}}\,^{d]\dot\b}
+m\Hat{\bf\J}_{\a\b\dot\a}=0
\eea

In summary, the fermionic sector of (\ref{finalaction}) describes two propagating
fermions.  $\Hat {\bf\J} _{(\a\b)\dot\b}$ satisfies (\ref{Dirac32}) and (\ref{nodiv}),
thus, forming a spin-${\fracm 32}$ representation, and $\j^{\a\dot\b}_{~~\a}$ forms a
spin-${\fracm 12}$ representation.  All other fermionic components of $V_\a$ vanish
except, for $\l_\a$ which is proportional to $\j^{\a\dot\b}_{~~\a}$.

%%%%%%%%%%%%%%%%%%%%%%%%%%%%%%%%%%%%%%%%%%%%%%%
\noindent
{\bf 5}.~~~~The analysis of the massless limit of the action (\ref{finalaction}) is quite
interesting.  As it turns out, this limit describes a reducible representation of the
massless Poincar\'e supergroup for arbitrary values of $\g$.  But, in the special case when
$\g={\fracm 14}$, the representation is irreducible and corresponds to the standard
gravitino multiplet.

The irreducible representation corresponding to the gravitino multiplet is given by the
chiral field strength $W_{\a\b}:=\Bar D^2 D_{(\a}V_{\b)}$  \cite{SJGWS} and it's equation of
motion:
\bea
\label{eomw}
D^\a W_{\a\b}=0
\eea
It is invariant under the gauge transformations:
\bea
\label{gauge1}
\d V_\a = \L_\a + i D_\a K\cr
K=\Bar K~~~~~\Bar D_{\dot\a}\L_\a=0
\eea
We will show that for arbitrary values of $\g$ this multiplet is contained in the massless
limit of (\ref{finalaction}). The gravitino multiplet is described in detail in
\cite{SJGWS}, \cite{Wit}  and \cite{Fradkin}.  The connection between the different
formalisms is stated rather concisely in \cite{Buch1}.

Setting $m=0$ leads us to the following action:
\bea
\label{maslesaction}
{\cal S}_{m=0}[V_\a,\F]=\int d^8z\Big \{
-V^\a\Bar D_{\dot\a}D_\a\Bar V^{\dot\a}
+|\g|V^\a\Bar D^2V_\a
+|\g|\Bar V_{\dot\a} D^2\Bar V^{\dot\a}\cr
-{\fracm 12}\F\Bar\F
-{\fracm 12}\F D^\a V_\a-{\fracm 12}\Bar \F \Bar D_{\dot\a} \Bar V^{\dot\a}
 \Big\}
\eea
Here, we have absorbed the phase of $\g=|\g|e^{i\f}$ by making the following field
redefinitions:
\bea
V_\a \rightarrow e^{-{\fracm i2}\f}V_\a~~~~~\F \rightarrow e^{{\fracm i2}\f}\F
\eea
This is only possible now because $m=0$.

To see what gauge symmetries are present we look for the massless representations implied
by the equations of motion.  The equations of motion for $V_\a$ and $\F$ are:
\bea
\label{eom1}
E'_\a := -\Bar D_{\dot\a} D_\a \Bar V^{\dot\a}
+2|\g |\Bar D^2 V_\a
+{\fracm 12}D_\a \F =0
\eea
\bea
\label{eom3}
{\fracm 18}D^2\F+{\fracm 18}D^2\Bar D_{\dot\a} \Bar V^{\dot\a}=0
\eea
Taking the divergence of (\ref{eom1}), i.e. $\pa_{\dot\a}^{~\a}E'_\a$, and substituting for
$\F$ using (\ref{eom3}) we find:
\bea
\label{irrepnomas}
\pa_{\dot\b}^{~\a}E'_\a=-{\fracm i8}\Bar D^{\dot\a} \Bar W_{(\dot\a\dot\b)}
+{\fracm i2}(4|\g |^2-{\fracm 14})\Bar D^2 D^2 \Bar V_{\dot\b}=0
\eea
Note that if $|\g |= {\fracm 14}$ we have the equation of motion for $W_{\a\b}$,
(\ref{eomw}).  At this point we turn our attention to the following projection of (\ref{eom1}):
\bea
\label{thebomb}
D^\a{\Bar D^2D^2\over 16\Box}E'_\a={\fracm 12}\Bar D_{\dot\a}D^2\Bar V^{\dot\a}
+2|\g |D^{\a}\Bar D^2 V_\a=0
\eea
This equation is of the form, $4|\g |H=-\Bar H$, which requires that either $|\g |={\fracm
14}$ or $D^\a\Bar D^2 V_\a=0$.  Since $D^\a\Bar D^2 V_\a=0\Rightarrow D^2\Bar D^2 V_\a=0$,
we see that (\ref{irrepnomas}) is equivalent to the equation of motion of $W_{\a\b}$,
(\ref{eomw}), for any value of $|\g|$, meaning that $W_{\a\b}$ propagates
for arbitrary $\g$.

When $|\g |={\fracm 14}$ the action (\ref{maslesaction}) is equivalent to the descriptions
of the gravitino multiplet given in the literature.\footnote{ We point out that there
exists another off-shell formulation of the gravitino multiplet in the literature
(see second paper in \cite{Kuzenko} and \cite{Buch1}). Such a formulation is given in
terms of a constrained complex linear transverse vector superfield or equivalently in
terms of spin-tensor superfield $\Psi_{\a\b\dot\a}$.}  This was shown in detail in
\cite{Buch1}.  In the case where $|\g |\not= {\fracm 14}$, equation (\ref{thebomb}) implies
the equation of motion for a superhelicity-${\fracm 12}$ representation.  That is, $D^\a \Bar
D^2V_\a=0$ means that the chiral field strength $\O_\a=\Bar D^2V_\a$ is an on-shell
representation of the massless Poincar\'e supergroup.  The gauge transformations that leave
both $W_{\a\b}$ and $\O_\a$ invariant are:
\bea
\d V_\a = \L_\a + i D_\a K\cr
\Bar D_{\dot\a}K=0~~~~~\Bar D_{\dot\a}\L_\a=0
\eea
In comparison to the gauge transformations that leave only $W_{\a\b}$ invariant,
(\ref{gauge1}), there is less gauge freedom here since $K$ is a chiral field.  

We have shown that the $m=0$ limit of the superspin-1 model (\ref{finalaction}) generically forms a
reducible representation of the massless Poincar\'e supergroup.  This massless model
(\ref{maslesaction}) contains superhelicity-$1$ and superhelicity-${\fracm 12}$ irreducible
representations.  Furthermore, if we set $\g={\fracm 14}$, this model contains only the
superhelicity-$1$ state, and is equivalent to the gravitino multiplet.  

%%%%%%%%%%%%%%%%%%%%%%%%%%%%%%%%%%%%%%%%%%%%%%%%%%%%%%%%%%%%%%%%%%%%%%%%%%%%%
\noindent{\bf 6.}  To summarize, we have presented a new $4D, N=1$ supersymmetric
model which describes propagating spin-$3/2$, spin-$1$ and spin-$1/2$ massive fields. The
model is completely formulated in terms of a spinor superfield $V_{\alpha}$ and chiral scalar
superfield ${\Phi}$. The superfield $V_{\alpha}$ is propagating and carries the superspin-$1$
massive irreducible representation of the Poincar\'e supergroup. The chiral superfield
${\Phi}$ is auxiliary and its role is to ensure the existence of a Lagrangian formulation
that is compatable with the conditions defining the massive irredicible representation of
superspin-$1$. The corresponding superfield action is given by equation
(\ref{finalaction}).

Equation (\ref{finalaction}) actually represents a two parametric family of actions that
all lead to the same on-shell dynamics. In the massless limit, this
two parametric family of actions becomes a one parametric family of actions
(\ref{maslesaction}). These massless models describe propagating helicity-$3/2$,
helicity-$1/2$ and two helicity-$1$ fields.

In terms of massive theories with arbitrarily high integer superspin, the superspin-1 theory
is the simplest.  In the same sense the model constructed in ref \cite{Buch2} is simplest
Lagrangian for half-integer higher superspin massive fields.  We believe that these two
models can be considered as the basis for constructing Lagrangian superfield models with
arbitrary integer and half-integer superspins.

\noindent
Added Note:

Near the completion of our work, we noted the appearance of a work by
Engquist, Sezgin and Sundell  \cite{ESS}.  Their work seems closely related to both the
topic of this letter as well as to some of the research that has appeared in
\cite{Kuzenko} which presented results on AdS geometry, SUSY and higher spin multiplets.
\vspace{0.5cm}
${~~~}$\newline
${~~~~~~~~~}$``{\it {The art of doing mathematics consists in finding that
special case which \newline ${~~~~~~~~~}$ contains all the germs of generality.}}"\newline
${~~~~~~~~~~~}$ --D. Hilbert

${~~~}$

%%%%%%%%%%%%%%%%%%%%%%%%%%%%%%%%%%%%%%%%%%%%%%%%%%%%%%%%%%%%%%%%%%%%%%%%
\noindent{\bf Acknowledgements}

\noindent I.L.B. is grateful to INTAS grant, project No 991-590, DFG grant, project 
No 436 RUS 113/669 and RFBR-DFG grant, project No 02-0204002 for partial
support.  Three of the authors, S.J.G., W.D.L. and J.A.P., would like to
acknowledge, John H. Schwarz and H. Tuck for the hospitality extended during their
visit to the California Institute of Technology, where most of this research was
undertaken.  Additionally, S.J.G. wishes to recognize the support rendered by
the Caltech administration during this visit.  The authors are also grateful to
S.M. Kuzenko for his useful comments.

%%%%%%%%%%%%%%%%%%%%%%%%%%%%%%%%%%%%%%%%%%%%%%%


\begin{thebibliography}{66}
\bibitem{Singh} L.P.S. Singh and C.R. Hagen, Phys.Rev. D9 (1974) 898; 919.

\bibitem{Kuzenko} S.M. Kuzenko, V.V. Postnikov, A.G. Sibiryakov, JETP Lett. 57 (1993)
534; S.M. Kuzenko and A.G. Sibiryakov, JETP Lett. 57 (1993) 539; S.M.
Kuzenko and A.G. Sibiryakov, Phys. Atom. Nuclei 57 (1994) 1257; S.J.
Gates, S.M. Kuzenko, A.G. Sibiryakov, Phys.Lett. B394 (1997) 343, hep-th/9611193; B412
(1997) 59, hep-th/9609141.

\bibitem{SJGWS} S. J. Gates, Jr. and W. Siegel, Nucl. Phys. {\bf B164},
484, 1980.

\bibitem{Buch1} I.L. Buchbinder and S.M. Kuzenko, {\it Ideas and Methods of Supersymmetry
and Supergravity}, IOP Publ., Bristol and Philadelphia, 1995, Revised
Edition 1998.

\bibitem{Buch2} I.L. Buchbinder, S.James Gates, Jr, W.D. Linch, III, J. Phillips,
Phys.Lett. B535 (2002) 280, hep-th/0201096.

\bibitem{Fierz} M. Fierz and W. Pauli, Proc. R. Soc. A173 (1939) 211.

\bibitem{Zino1}Yu. M. Zinoviev, ``Massive Spin-2 Supermultiplets", hep-th/0206209.

\bibitem{Rarita} W. Rarita and J. Schwinger, Phys.Rev. 60 (1941) 61.

\bibitem{Biswas} T. Biswas and W. Siegel, JHEP 0207 (2002) 005, hep-th/0203115.

\bibitem{Lopatin} V.E. Lopatin and M.A. Vasiliev, Mod.Phys.Lett. A3 (1988) 257; M.A.
Vasiliev, Nucl.Phys. B301 (1988) 26; R.R. Metsaev, Phys.Lett. B354
(1995) 295; L. Brink, R.Metsaev, M. Vasiliev, Nucl.Phys. B586 (2000)
183, hep-th/0005136; I.L. Buchbinder, A. Pashnev, M. Tsulaia, Phys.Lett. B523 (2001)
1853, hep-th/0109067; Proc. XVI Max Born Symposium ``Supersymmetries and Quantum
Symmetries", Karpacz, Poland, 21-25 September 2001, Dubna 2002, 3, hep-th/0206026.

\bibitem{Berk} N. Berkovits and M.M. Leite, Phys.Lett. B415 (1997) 295, hep-th/9709148; B452
(1998) 38, hep-th/9812153; N. Berkovits and O. Chandia, ``Massive
Superstring Vertex Operator in D=10 Superspace", hep-th/0204121.

\bibitem{Buch3} I.L. Buchbinder, V.A. Krykhtin, V.D. Pershin, Phys.Lett. B466
(1999) 216, hep-th/9908028; I.L. Buchbinder, D.M. Gitman, V.A. Krykhtin, V.D. Pershin,
Nucl.Phys. B584 (2000) 615, hep-th/9910188; I.L. Buchbinder, D.M. Gitman, V.D. Pershin,
Phys.Lett. B492 (2000) 161, hep-th/0006144; I.L. Buchbinder and V.D. Pershin, in volume
``Geometrical aspects of Quantum Fields", World Scientific, 2001, 11, hep-th/0009026.

\bibitem{Deser} S. Deser and A. Waldron, Phys.Rev.Lett. 87 (2001) 031602, hep-th/0102166;
Phys.Lett. B508 (2001) 347, hep-th/0103255; B513 (2001) 127, hep-th/0105181; Nucl.Phys. B607
(2001) 577, hep-th/0103198.

\bibitem{Dolan} L. Dolan, C.R. Nappi, E. Witten, JHEP 0110 (2001) 016, hep-th/0109096.

\bibitem{Zino2}Yu. M. Zinoviev, ``On Massive Higher Spin Particles in (A)dS",
hep-th/0108192.

\bibitem{Francia}D. Francia and A. Sagnotti, ``Free Geometrical Equations for Higher
Spins", hep-th/0207002.

\bibitem{Segal}A.Y. Segal, ``Conformal higher spin theory", hep-th/0207212.

\bibitem{Siegel1} W. Siegel and S. J. Gates, Jr., Nucl. Phys. {\bf B189}, 295, 1981.

\bibitem{Gates1} S. J. Gates, Jr., M. T. Grisaru, M. Rocek, W. Siegel, {\it Superspace},
Benjamin/Cummings Publishing Co., Inc. (Reading, MA), 1983, hep-th/0108200.


\bibitem{Wit} B. de Wit and J. W. van Holten, Nucl. Phys. {\bf B155}, 530, 1979.

\bibitem{Fradkin} E. S. Fradkin and M. A. Vasiliev, Nuovo Cimento Lett. {\bf 25}, 79,
1979.

\bibitem{ESS} J. Engquist, E. Sezgin, P. Sundell, ``On N=1,2,4 Higher
Spin Gauge Theories in Four Dimensions," hep-th/0207101.


\end{thebibliography}
\end{document}